\documentclass[preprint,showkeys]{revtex4}
\usepackage{graphicx}
\usepackage{epsfig}
\usepackage{amssymb}
\usepackage{color}
\usepackage{amsmath}
\usepackage{lipsum}
\usepackage[font={small}]{caption}
\usepackage{graphics}
\usepackage{tensor}
\usepackage{filecontents}
\usepackage{epsf}
\usepackage{psfrag}
\usepackage{epstopdf}
\usepackage{graphicx}
\usepackage{caption}
\usepackage{epstopdf}
\usepackage{epsfig}

\begin{document}

\title{Non-Extensive Entropy and Power-Law Inflation: Implications for Observations}
\author{ A.Khodam-Mohammadi \footnote{Email:khodam@basu.ac.ir}} 
\affiliation{Department
of Physics, Faculty of Science, Bu-Ali Sina University, Hamedan
65178}

\begin{abstract}
This study explores the interaction between non-extensive entropic FLRW cosmology and the power-law inflationary model, with a focus on the overlap between the scalar spectral index `$n_s$' and the tensor-to-scalar ratio `$r$'. Based on a conjecture that non-extensive entropy alters the energy-momentum content of the cosmic fluid, the analysis examines how these overlaps shift with different model parameters and compares the findings to those from Bekenstein-Hawking (BH) entropic cosmology. The study highlights the impact of Tsallis, R\'{e}nyi, and Sharma-Mittal entropies, uncovering a significant correlation between `$n_s$' and `$r$' that suggests a deeper connection in power-law inflationary dynamics. The results demonstrate that non-extensive entropies not only enable viable inflation with a graceful exit but also address limitations inherent in the standard BH entropic framework, emphasizing the importance of precise parameter estimation. Specifically, Tsallis entropy allows for power-law inflation with $n = 1$ to $n = 2$ in alignment with Planck 2018 data. Moreover, the $\alpha$ parameter in R\'{e}nyi and Sharma-Mittal entropy models must be extremely small ($\alpha \leq 10^{-8}$ in Planck mass units) to achieve successful power-law inflation with an e-folding number around 55-65, suggesting a unified thermodynamic perspective in cosmological studies.\end{abstract}


\maketitle 

\section{Introduction}
The application of event horizons to black holes within the Bekenstein-Hawking (BH) and other non-extensive entropy framework, along with the study of their thermodynamics, has inspired cosmologists to extend this approach to cosmological apparent horizons \cite{Bekenstein:1973ur, Hawking:1975vcx, Jacobson:1995ab, Hayward:1997jp, Padmanabhan:2003gd, Padmanabhan:2009vy, Padmanabhan:2013nxa, Sanchez:2022xfh, Sheykhi_2018, Odintsov:2022qnn}. This methodology aims to address key issues in cosmology, including inflationary cosmology  \cite{Odintsov_2023, Odintsov:2023rqf, Lambiase:2023ryq, Teimoori:2023hpv, Luciano:2023roh}, the dark energy problem  \cite{Brevik:2024nzf, Sheykhi_2023, P:2022amn, Nojiri_2022, Manoharan:2022qll, Di_Gennaro_2022, Bhattacharjee_2021, Moradpour:2020dfm, DAgostino:2019wko, Saridakis_2018}, and the Hubble tension problem  \cite{Hern_ndez_Almada_2022, Asghari:2021bqa, Dabrowski:2020atl}. As a result, there has been a surge in applying non-extensive entropies to the Universe's apparent horizon. Beyond the well known BH entropy, Tsallis entropy \cite{Tsallis:1987eu} has been extensively studied both theoretically and cosmologically. Other forms of entropy, such as R\'{e}nyi \cite{Renyi:1960}, Sharma-Mittal \cite{SayahianJahromi:2018irq}, Barrow \cite{Barrow:2020tzx}, Kaniadakis \cite{Kaniadakis:2005zk,Drepanou:2021jiv}, and loop quantum gravity  \cite{Majhi:2017zao,Liu:2021dvj} entropies, have also been rigorously studied.

Recently, two distinct methods have been proposed for investigating FLRW cosmology by incorporating generalized entropies. The first method focuses on modifying FRW cosmology without altering the cosmic fluid or the continuity equation  \cite{Sheykhi_2018, Odintsov:2022qnn, Nojiri_2022}. This is done by directly adjusting the Friedmann equations based on generalized entropy considerations, facilitating the exploration of dark energy models and early universe inflation. In this approach any exact solution of Friedmann equations are not obtained and consequently results are given by any approximately relations.  

Conversely, a new method, developed by myself and a colleague  \cite{Khodam_Mohammadi_2023}, incorporates the effects of non-extensive entropy by introducing a conjecture that say ``\textit{By generalizing entropy in an FLRW cosmology, the energy-momentum density tensor of the cosmic fluid is modified in such a way that a common correction function multiplies on both the energy density and pressure terms. This ensures that the perfectness property and the EoS parameter of the cosmic fluid remain unchanged.}''. This modification results in changes to the continuity equation and subsequently leads to revised Friedmann equations. The strength of this approach lies in its precise calculation of the corrective factor and the exact form of the modified Friedmann equations, whereas the previous method.

Additionally, this new approach satisfies the first and second laws of thermodynamics and examines the effects of the corrective factor on cosmographic quantities in a non-flat FLRW universe  \cite{Khodam_Mohammadi_2023}.
In this study, we focus on inflationary cosmology incorporating non-extensive entropy, specifically within a framework where the matter fields inside the apparent horizon are obtained from a scalar field characterized by a power-law potential. In the absence of such a matter fields, entropic cosmology typically evolves into a de Sitter space-time, analogous to an eternal inflationary state with no exit. Therefore, to achieve a viable inflationary model, it is essential to include appropriate matter fields within the horizon \cite{Odintsov:2023rqf}.

The motivation for adopting a power-law scalar potential originates from the inadequacy of the simple $\phi^2$ potential in standard scalar field cosmology, which has been ruled out by the recent Planck 2018 data due to its inconsistency with observed inflationary parameters  \cite{Odintsov:2023rqf}.

However, the scenario changes significantly when the horizon entropy adopts a non-extensive form, such as Tsallis, R\'{e}nyi,  Sharma-Mittal or in general four-parameters entropies  \cite{Odintsov:2023rqf}. In such cases, the modifications to the cosmic fluid resulting from the horizon entropy become crucial, as they significantly influence the universe's evolution by introducing corrections to the Friedmann equations  \cite{Khodam_Mohammadi_2023}. Motivated by these considerations, this paper aims to investigate the feasibility and implications of the $\phi^n$-matter field potential within the context of entropic inflation, where the horizon entropy is governed by any form of non-extensive entropy. This approach allows for a more comprehensive examination of how generalized entropy forms can impact inflationary dynamics and the broader cosmological model.

This paper is organized as follows: Section \ref{Sec2} provides a brief review of the modified Friedmann equations within the framework of generalized entropy models. In section \ref{Sec3}, I investigate the behavior of the $\phi^n$-matter field inflationary cosmology when the horizon entropy is altered by some classes of non-extensive entropy such as Tsallis, R\'{e}nyi and Sharma-Mittal entropies. Furthermore in section \ref{Sec4}, these results are compared and discussed with the standard cosmology context governed by BH entropy. Finally, section \ref{Sec5} offers concluding remarks and a summary of our key findings.
 
Throughout this article, we employ Planck mass units where $c = G = \hbar = 1$. An over-dot will denote differentiation with respect to time, while a prime will signify differentiation with respect to the variable itself.

\section{Modification of FLRW cosmology from a general entropy}\label{Sec2}
This section provides a brief review of the new approach to general entropy FLRW cosmology, as presented in \cite{Khodam_Mohammadi_2023}.
 
In an isotropic FLRW metric,
\begin{equation}
ds^2=h_{ab}dx^a dx^b+R^2(d\theta^2+sin^2\theta d\phi^2),
\end{equation}
where $x^0=t,~ x^1=r,~ R=a(t)r$ and $h_{a b}=diag\Big{(}-1,~a(t)^2/(1-kr^2)\Big{)}$ with its determinant, `$h$', the BH entropy, apparent horizon radius, surface gravity, and temperature are given by  \cite{Sanchez:2022xfh,Cai:2005ra,Binetruy:2014ela,Hayward:1997jp}
\begin{eqnarray}
S_{BH}&=&\frac{A}{4}=\pi R_h^2=\frac{3}{8\rho} \\
R_h^2&=&\frac{1}{(H^2+\frac{k}{a^2})} \\
\kappa&=&\frac{1}{2\sqrt{-h}}\frac{\partial}{\partial x^a}(\sqrt{-h}h^{a b}\frac{\partial R_h}{\partial x^b})\\
T_h&=&\frac{|\kappa|}{2\pi}=\frac{1}{2\pi R_h}|1-\frac{\dot{R_h}}{2 H R_h}|.
\end{eqnarray}

Considering the modified energy-momentum tensor of a perfect fluid
\begin{equation}
T^\mu_\nu=diag(-\rho f(\rho), p f(\rho),p f(\rho),p f(\rho)), \label{new-fluid}
\end{equation} 
with the equation of state parameter (EoS) $w=p/\rho$, the conservation law $T^{\mu\nu}_{;\nu} =0$, leads to
\begin{equation}
\dot{\rho}[1+\rho\frac{f^\prime(\rho)}{f(\rho)}]+3H(\rho+p)=0.\label{consf}
\end{equation}

By considering the new perfect fluid (\ref{new-fluid}), the modified Friedmann equations become
\begin{eqnarray}
H^2+\frac{k}{a^2}&=&\frac{8 \pi}{3} \rho f(\rho)+\frac{\Lambda}{3} \notag \\
\dot{H}-\frac{k}{a^2}&=&-4 \pi (\rho+p) f(\rho).\label{FR}
\end{eqnarray}
where $\Lambda$ known as the cosmological constant, appears as an integration constant.

Using the modified Friedmann equations (\ref{FR}) and the first law of thermodynamics, $dE=-\delta Q+W d V$, where $W=-1/2 T^{a b}h_{a b}$ is the work density and $\delta Q=T d S$, and also using
\begin{equation}
E=\frac{4}{3}\pi R^3_h \rho f(\rho),~~ V=\frac{4}{3}\pi R^3_h,~~ W=-\frac{1}{2}(p-\rho) f(\rho),
\end{equation}
we obtain the following relation
\begin{equation}
\tilde{S}=S_g(S), \label{G-entropy}
\end{equation}
where $S_g$ is any non-extensive entropy. Here, $S=\frac{\pi r^2_h}{4}=\frac{3}{8\rho}$ represents the standard BH entropy, and $\tilde{S}=\frac{3}{8\rho f(\rho)}$ represents the BH entropy in the context of new cosmic fluid (\ref{new-fluid}). The relation (\ref{G-entropy}) provides an innovative method for determining $f(\rho)$ in any non-extensive entropic model.

Table \ref{chart} lists some correction functions $f(\rho)$ corresponding to various entropy functions $S_g$.
 \begin{table*}[h]
	\begin{center}
		\begin{tabular}{|r|c|c|c|l|}
			\hline
			
			&Entropy & $S_g$ & $f(\rho)$ \\
			\hline
			1 &Tsallis & $S_0(\dfrac{S}{S_0})^\delta$ & $(\dfrac{8}{3}S_0\rho)^{\delta-1}$ \\
			\hline
			2 &R\'{e}nyi & $ \dfrac{1}{\alpha}\ln(1+\alpha S)$ & $\dfrac{3\alpha}{8\rho}[\ln(1+\dfrac{3\alpha}{8\rho})]^{-1} $ \\
			\hline
			3 &Sharma-Mittal &$\dfrac{1}{R}[(1+\alpha S)^{R/\alpha}-1]$ & $\dfrac{3R}{8\rho}[(1+\dfrac{3\alpha}{8\rho})^{R/\alpha}-1]^{-1} $ \\
			\hline
     		\end{tabular}
	\end{center}
\caption{Reconstruction of $f(\rho)$ for various entropy functions  \cite{Khodam_Mohammadi_2023}}. \label{chart}
\end{table*}
\section{Entropic Inflation with $\phi^n$-Matter Field}\label{Sec3}
\subsection{Bekenstein-Hawking entropy}\label{subsecA}
First of all, in a standard flat FLRW cosmology, by considering $ S_g = S$, the standard Friedmann equations and continuity equation by considering following scalar matter field $\rho=\dot{\phi}^2/2+V(\phi)$ and the pressure $p=\dot{\phi}^2/2-V(\phi)$, gives \cite{Odintsov:2023rqf}

\begin{eqnarray}
&&H^2 = \frac{8 \pi}{3} \left( \frac{\dot{\phi}^2}{2} + V(\phi) \right), \\
&&\dot{H}= -4 \pi \dot{\phi}^2, \\
&&\ddot{\phi} + 3 H \dot{\phi} + \partial_{\phi} V = 0.
\end{eqnarray}

Using the slow-roll approximation, where $ V(\phi) \gg \dot{\phi}^2 $ and $ \ddot{\phi} \ll H \dot{\phi} $), and with $ \dot{\phi} \approx -\frac{\partial_{\phi} V}{3H}$, the slow-roll parameters are given by:

\begin{eqnarray}
\epsilon &=& -\frac{\dot{H}}{H^2} = \frac{3 \dot{\phi}^2}{2 V(\phi)}, \notag\\
\eta &=& \frac{\ddot{\phi}}{H \dot{\phi}} = -\sqrt{\frac{3}{8 \pi V(\phi)}} \frac{\ddot{\phi}}{\dot{\phi}}.\label{epseta}
\end{eqnarray}

Assuming the power-law scalar potential $ V(\phi) = V_0 \phi^n $, the slow-roll parameters become

\begin{eqnarray}
\epsilon &=& \frac{n^2}{16 \pi \phi^2}, \\
\eta &=& \frac{n (n - 2)}{16 \pi \phi^2}. \label{epseta1}
\end{eqnarray}

As the scalar field rolls down, the parameter $\epsilon$ increases up to unity and an inflation with a graceful exit is achieved. At this point, inflation ends at $ \phi_f = \frac{n}{\sqrt{16 \pi}} $, and the total number of e-folds at this time becomes:

\begin{equation}
N_f = \int_{\phi_c}^{\phi_f} \frac{H}{\dot{\phi}} d\phi \approx \frac{4 \pi}{n} \left( \phi_c^2 - \frac{n^2}{16 \pi} \right), \label{NF}
\end{equation}
where $\phi_c$ is the scalar field value at the time of horizon crossing of the CMB mode. By solving for $\phi_c$ from Eq. (\ref{NF}) and substituting it into Eq. (\ref{epseta1}), one find:

\begin{eqnarray}
\epsilon_c &=& \frac{n}{4 N_f + n}, \label{epseta2}\\
\eta_c &=& \frac{n - 2}{4 N_f + n}. 
\end{eqnarray}

Finally, the spectral index for primordial curvature perturbation  \cite{Odintsov:2023rqf} given by $n_s=1-6\epsilon_c+2\eta_c$ and the tensor-to-scalar ratio $r=16\epsilon_c$ at the horizon crossing are 

\begin{eqnarray}
n_s &=& 1 - \frac{4 (n + 1)}{4 N_f + n}, \\
r &=& \frac{16 n}{4 N_f + n}. \label{ns-r}
\end{eqnarray}
The key question is whether the entropy modifications influence the spectral index $n_s$ and the tensor-to-scalar ratio $r$. We know these parameters are derived from the slow-roll parameters $\epsilon$ and $\eta$, which are in turn influenced by the first and second time derivatives of the scalar field $\phi$. It is natural that with the modification of Friedman's equations due to the choice of specific non-extensive entropic model, the potential function and as a result the parameters of slow roll change, but the relationship of $n_s$ and $r$ parameters in terms of $\epsilon$ and $\eta$ does not change (see the appendix \ref{app}).
 
In the next stage, I am interested in models whose apparent horizon entropy belongs to a class of non-extensive entropies. Among the most well-known of these are the Tsallis entropy, followed by examining two other prominent entropy models, namely R\'{e}nyi and Sharma-Mittal entropies.
\subsection{Tsallis entropy} \label{Tsal2}
In the context of Tsallis entropy, the apparent horizon entropy is expressed as
\begin{equation}
S_g=S_0(\frac{S}{S_0})^\delta.
\end{equation}
The Friedmann equations and continuity equation (\ref{consf}, \ref{FR}) in a flat FLRW cosmology in the absence of $\Lambda$ are modified as
\begin{eqnarray}
&&H^2 = \frac{\pi}{S_0} \left( \frac{8}{3} S_0 \rho  \right)^\delta, \\
&&\dot{H}= -4 \pi (\rho+p) \left( \frac{8}{3} S_0 \rho  \right)^{\delta-1}, \\
&& \dot{\rho}~\delta+ 3 H (\rho+p) = 0.
\end{eqnarray}
Using $V(\phi)=V_0 \phi^n$, for matter field under the assumption of slow roll inflation, above equations are rewritten as
\begin{eqnarray}
H^2 &=& \frac{\pi}{S_0} \left( \frac{8}{3} S_0 V_0   \right)^\delta\phi^{n\delta}, \\
\dot{H} &=&-\frac{1}{6} V_0 n^2 \delta^2 \phi^{n-2}, \\
\dot{\phi} &=&-\frac{1}{3} n \delta V_0 \left(\frac{8}{3}S_0 V_0\right)^{-\frac{\delta}{2}} \phi^{2n-n \delta -2}.
\end{eqnarray}  
After some calculations, the slow roll parameters, $\epsilon$ and $\eta$, analogous to the previous relations (\ref{epseta}), are
\begin{eqnarray}
\epsilon &=& \frac{n^2 \delta^2}{16 \pi}\left(\frac{8 }{3}S_0 V_0\right)^{1-\delta}\tilde{\phi}^{-1}, \\
\eta &=& \frac{n(n-2) \delta}{16 \pi}\left(\frac{8 }{3}S_0 V_0\right)^{1-\delta}\tilde{\phi}^{-1}, \\
\frac{\epsilon}{\eta} &=& \frac{n-2}{n \delta}, \label{epseta-tsal}
\end{eqnarray} 
where $\tilde{\phi}=\phi^{n \delta - n+2}$. Inflation ends when 
\begin{equation}
\tilde{\phi_f}=\frac{1}{16\pi}\frac{n^2 \delta^2} {\left( \frac{8}{3} S0 V0 \right) ^{\delta-1}},
\end{equation}
yielding the number of e-folds at the end of inflation as:
\begin{equation}
N_f=\frac{8\pi}{n \delta(n\delta-n+2)}\left(\frac{8}{3} S0 V0 \right)^{\delta-1}\left(\frac{1}{16\pi}\frac{n^2 \delta^2} {\left( \frac{8}{3} S0 V0 \right) ^{\delta-1}}-\tilde{\phi_c}\right). \label{Nf-tsal}
\end{equation}
Solving $\tilde{\phi_c}$ from (\ref{Nf-tsal}) and substituting in (\ref{epseta-tsal}), the slow roll parameters and consequently the power spectrum index $n_s$ and tensor to scalar ratio $r$ at the horizon crossing are calculated as
\begin{eqnarray}
\epsilon_c &=& \frac{n \delta}{n \delta+(2 n \delta-2 n +4)N_f},\notag \\
\eta_c &=& \frac{n -2}{ n \delta+(2 n \delta-2 n +4)N_f} \notag \\
n_s &=& 1-\frac{2(3 n \delta - n +2)}{n \delta+(2 n \delta-2 n +4)N_f}, \notag \\
r &=& \frac{16 n \delta}{n \delta+(2 n \delta-2 n +4)N_f}. \label{nsr-tsal}
\end{eqnarray}  
It is noteworthy that in limiting case, $\delta=1$, all the above equations reduce to those for standard BH entropy, Eqs. (\ref{epseta2}-\ref{ns-r}). This fact demonstrates the accuracy of precise calculations of the quantities mentioned in this approach. 
\subsection{R\'{e}nyi entropy} \label{Ren}
The entropy associated with the apparent horizon in the R\'{e}nyi entropy framework, as shown in the second row of table (\ref{chart}), is given by
\begin{equation}
S_g=\frac{1}{\alpha}\ln(1+\alpha S),
\end{equation}
where $\alpha$ is a model parameter of dimension of the inverse of entropy. 
By using the corresponding function  $f(\rho)$ and considering the power-law scalar potential for the matter field, after applying the slow roll considerations like previous section, Friedmann equations can be rewritten as
\begin{eqnarray}
H^2 &=& \frac{8\pi}{3}\frac{V_0 \Gamma}{ln(1+\Gamma {\varphi})}, \\
\dot{H} &=&-\frac{1}{6}\frac{ V_0 n^2 \Gamma^2 }{ln(1+\Gamma \varphi)^2(1+\Gamma \varphi)^2}\varphi^{1+2/n}, \\
\dot{\phi} &=&-\sqrt{\frac{8\Gamma V_0}{3 \pi}}\frac{n }{(1+\Gamma \varphi)\sqrt{ln(1+\Gamma \varphi)}}\varphi^{1/n}.
\end{eqnarray}  
Here $\Gamma=3 \alpha/(8V_0)$ is a new dimensionless parameter made from $\alpha~\&~V_0$ and also $\varphi=\phi^{-n}$.  
Using above equations, slow roll parameters are derived as
\begin{eqnarray}
\epsilon &=& \frac{1}{16\pi}\frac{n^2 \Gamma }{(1+\Gamma \varphi)^2 ln(1+\Gamma \varphi)}\varphi^{1+2/n}, \label{epsil-ren}\\
\eta &=& \frac{n}{16\pi}\frac{ln(1+\Gamma \varphi)[\Gamma  (n-2)-(n+2)\varphi^{-1}]+2 n \Gamma}{(1+\Gamma \varphi)^2 ln(1+\Gamma \varphi)}\varphi^{1+2/n}, \\
\frac{\epsilon}{\eta} &=&\frac{ n \Gamma}{ln(1+\Gamma \varphi)[\Gamma  (n-2)-(n+2)\varphi^{-1}]+2 n \Gamma}.
\end{eqnarray}  
It seems that the positivity of $\epsilon$ demands $\Gamma>0$. The number of e-fold at the end of inflation is calculated by
\begin{equation}
N_f=-\frac{4\pi }{n}\left\{\phi_f^2\left(1-\frac{2 \Gamma \phi_f^{-n}}{n-2}\right)-\phi_c^2\left(1-\frac{2 \Gamma \phi_c^{-n}}{n-2}\right)\right\}. \label{Nf-Renl}
\end{equation}
As we see, the Eq. (\ref{Nf-Renl}) diverges at $n=2$ and we must exclude it from the allowed values of $n$. The scalar field value $\phi_f$ at the end of inflation is determined by solving  $\epsilon=1$. Although according to (\ref{epsil-ren}), an exact analytical solution of this equation is not feasible, but by expanding $\epsilon$ to the first order in terms of $\Gamma$ and in the case of $n=1$, one of possible solution is given
\begin{equation}
\varphi_f=\frac{2}{9\Gamma}\left[(\mathcal{M}+\sqrt{\mathcal{M}^2-1})^{1/3}+(\mathcal{M}+\sqrt{\mathcal{M}^2-1})^{-1/3}+1\right], \label{phif-R}
\end{equation}
where $\mathcal{M}=(1-486\pi \Gamma^2)$. It must be mentioned that the parameter $\Gamma \ll 1 $. To prove that $\Gamma$ is very small during the inflation period in Planck mass units, consider that $\Gamma = \frac{3\alpha}{8V_0}$, where $\alpha$ is of the order of inverse entropy and $V_0$ is the order of energy density during inflation which is $\sim 10^{-8}M_P^4$ in Planck mass unit. Given that the entropy $S$ is inversely proportional to $H^2/\pi$ and the Hubble parameter $H^2$ is of the order $10^{-8} M_P^2$ during inflation, $\alpha$ becomes extremely small, $\alpha < 10^{-8}$. Specifically, for $\alpha \sim 10^{-13}$ up to $10^{-8}$, this results in $\Gamma \approx 10^{-5}$ up to $10^{-1}$ in Planck units. It ensuring $\Gamma/\phi^n < 1$ for typical values of $\phi \sim M_P=1$ and $n \approx 1$. This confirms that $\Gamma$ remains negligibly small, maintaining the validity of our assumptions throughout the inflation period \cite{mukhanov2005physical, tsallis2009introduction}. 

Substituting (\ref{phif-R}) into (\ref{Nf-Renl}), and expanding $\phi_c$ to the first order of $\Gamma$ we obtain:
\begin{equation}
\varphi_c=-2 \sqrt{\frac{\pi}{N_f}}+\frac{4\pi}{ N_f}\Gamma.
\end{equation} 
At last after calculating $\epsilon_c$ and $\eta_c$ at the horizon crossing, two quantities $n_s$ and $r$ are given by
\begin{eqnarray}
n_s &=&1-\frac{2}{N_f}+\frac{3}{2}\frac{\sqrt{\pi N_f}}{N_f^2}\Gamma,\notag \\
r &= &\frac{4}{N_f}(1-\frac{\sqrt{\pi N_f}}{N_f}\Gamma).\label{nsr-Ren}
\end{eqnarray}
\subsection{Sharma-Mittal entropy} \label{Sharm}
Similar to the previous case, the apparent horizon entropy is given in the third row of Table (\ref{chart}).
\begin{equation}
S_g=\dfrac{1}{R}[(1+\alpha S)^{R/\alpha}-1]
\end{equation}
where both model parameters $R$ and $\alpha$ have dimension of the inverse of entropy and similar to R\'{e}nyi entropy are exceedingly small.

Using the corresponding function  $f(\rho)$ and power-law scalar potential for the matter field and after applying the slow roll considerations, the Friedmann equations can be rewritten as
\begin{eqnarray}
H^2 &=& \frac{8\pi}{3}\frac{\gamma\Gamma V_0 }{(1+\Gamma \varphi)^\gamma-1}, \\
\dot{H} &=&-\frac{1}{6}\frac{ \gamma^2 \Gamma^2 V_0 n^2(1+\Gamma \varphi)^{2\gamma -2} }{[(1+\Gamma \varphi)^\gamma-1]^2}\varphi^{1+2/n}, \\
\dot{\phi} &=&-\sqrt{\frac{8\gamma \Gamma V_0}{3 \pi}}\frac{n(1+\Gamma \varphi)^{\gamma-1} }{\sqrt{(1+\Gamma \varphi)^\gamma-1}}\varphi^{1/n}.
\end{eqnarray}  
Here as before $\Gamma=3 \alpha/(8V_0)$ takes very small values, $\varphi=\phi^{-n}$ and new parameter $\gamma=R/\alpha$ has values of order unity or greater.  
Using above equations, slow roll parameters are given by
\begin{eqnarray}
\epsilon &=& \frac{1}{16\pi}\frac{\gamma \Gamma n^2(1+\Gamma \varphi)^{2\gamma-2}}{ (1+\Gamma \varphi)^\gamma-1}\varphi^{1+2/n}, \label{epsil-SM}\\
\eta &=& \frac{n}{16\pi}\frac{[(1+\Gamma \varphi)^\gamma-1][\Gamma  (n-2)-(n+2)\varphi^{-1}]+2 n \gamma \Gamma}{(1+\Gamma \varphi)^{2-\gamma} [(1+\Gamma \varphi)^\gamma-1]}\varphi^{1+2/n}.
\end{eqnarray}  
For $\epsilon >0$, it is required that $\gamma \Gamma>0$, implying both $\{\gamma$ and $\Gamma\}$  must be positive or negative. The number of e-folds at the end of inflation, to the first order in $\Gamma$ is
\begin{equation}
N_f=-\frac{4\pi }{n}\left\{\phi_f^2\left(1-\frac{2 \Gamma (1-\gamma) \phi_f^{-n}}{n-2}\right)-\phi_c^2\left(1-\frac{2 \Gamma (1-\gamma) \phi_c^{-n}}{n-2}\right)\right\}, \label{Nf-SM}
\end{equation}

As we see, similar to the previous case, Eq. (\ref{Nf-SM}) diverges at $n=2$ and we should exclude it from the allowed values of $n$. The quantity $\phi_f$ is used for scalar field at the end of inflation. This value is found by solving $\epsilon=1$. After expanding $\epsilon$ to the first order of $\Gamma$ and in the case of $n=1$, one possible solution is given by
\begin{equation}
\varphi_f=-\frac{2}{3\Gamma (\gamma-1)}+24\pi(\gamma-1)\Gamma. \label{phif-SM}
\end{equation}
Substituting (\ref{phif-SM}) into (\ref{Nf-SM}), the quantity $\phi_c$ after expanding its solution in first order of $\Gamma$ is obtained as
\begin{equation}
\varphi_c=-2 \sqrt{\frac{\pi}{N_f}}-\frac{4\pi}{N_f}(\gamma-1)\Gamma.
\end{equation} 
At last, after calculating $\epsilon_c$ and $\eta_c$ at the horizon crossing, two quantities $n_s$ and $r$ are given by
\begin{eqnarray}
n_s &=&1-\frac{2}{N_f}-\frac{3}{2}\frac{\sqrt{\pi N_f}}{N_f^2}(\gamma-1)\Gamma,\notag \\
r &= &\frac{4}{N_f}+4\frac{\sqrt{\pi N_f}}{N^2_f}(\gamma-1)\Gamma.\label{nsr-SM}
\end{eqnarray}
\subsection{General Four-Parameter Entropies and Other Entropy Forms}

When exploring four-parameter general entropy models, which can represent various non-extensive entropy types in specific limiting cases, several challenges arise. 

Firstly, calculations involving these four-parameter entropies, as well as other non-extensive entropy forms, tend to converge towards standard cosmological models that assume Bekenstein-Hawking (BH) entropy at the leading order of approximation. Consequently, the leading-order approximation does not provide any novel insights.

Secondly, accurate analysis requires calculations that go beyond the leading-order approximation. These start with approximate evaluations of the slow-roll parameters, $\epsilon$ and $\eta$, for specific integer values of the power-law index $n$. Subsequent approximations are then necessary for determining the scalar field at the end of inflation, followed by additional approximations for calculating the scalar field at the horizon crossing. This iterative process, required to ultimately calculate the scalar spectral index $n_s$ and the tensor-to-scalar ratio $r$, introduces significant cumulative errors, undermining the reliability of the results.

Moreover, selecting an integer value for the power-law index $n$ at the outset often leads to singular terms in subsequent calculations, creating mathematical inconsistencies that prevent further progress.

These issues suggest that investigating inflationary models using more complex entropies is fraught with computational difficulties, which effectively hinder practical analysis. However, this does not imply that such entropy models are without value; rather, they present challenges that require innovative approaches to overcome.

\section{Discussion and results} \label{Sec4}
In this section, we address a fundamental question: Can a power-law potential with a scalar field $\phi$ lead to successful inflation within a broader cosmological framework beyond the scope of black hole (BH) entropy? As discussed in  \cite{Odintsov:2023rqf} and based on the results of subsection (\ref{subsecA}), despite achieving a satisfactory number of e-folds $N \approx 60$, the $1\sigma$ confidence region for $n_s$ and $r$ only marginally overlaps around  $n \approx 1$ for $N_f\approx 60$. This overlap is statistically insignificant, rendering the model inconsistent with observations for other values of $n$. It's worth noting that the Planck 2018 data \cite{Odintsov:2023rqf} provides crucial constraints on the observational indices:
\begin{equation}
n_s = 0.9649 \pm 0.0042 \quad \text{and} \quad r < 0.064. \label{P2018}
\end{equation}
Consequently, it is evident that such an inflationary model based on a universe characterized by the entropy of the event horizon, as in the BH entropic paradigm, is inadequate. Hence, we turn our attention to exploring alternative non-extensive entropy frameworks, as discussed in Section (\ref{Sec3})
\subsection{Tsallis entropy}
\begin{figure}[t]
	\begin{center}
		\includegraphics[scale=0.48]{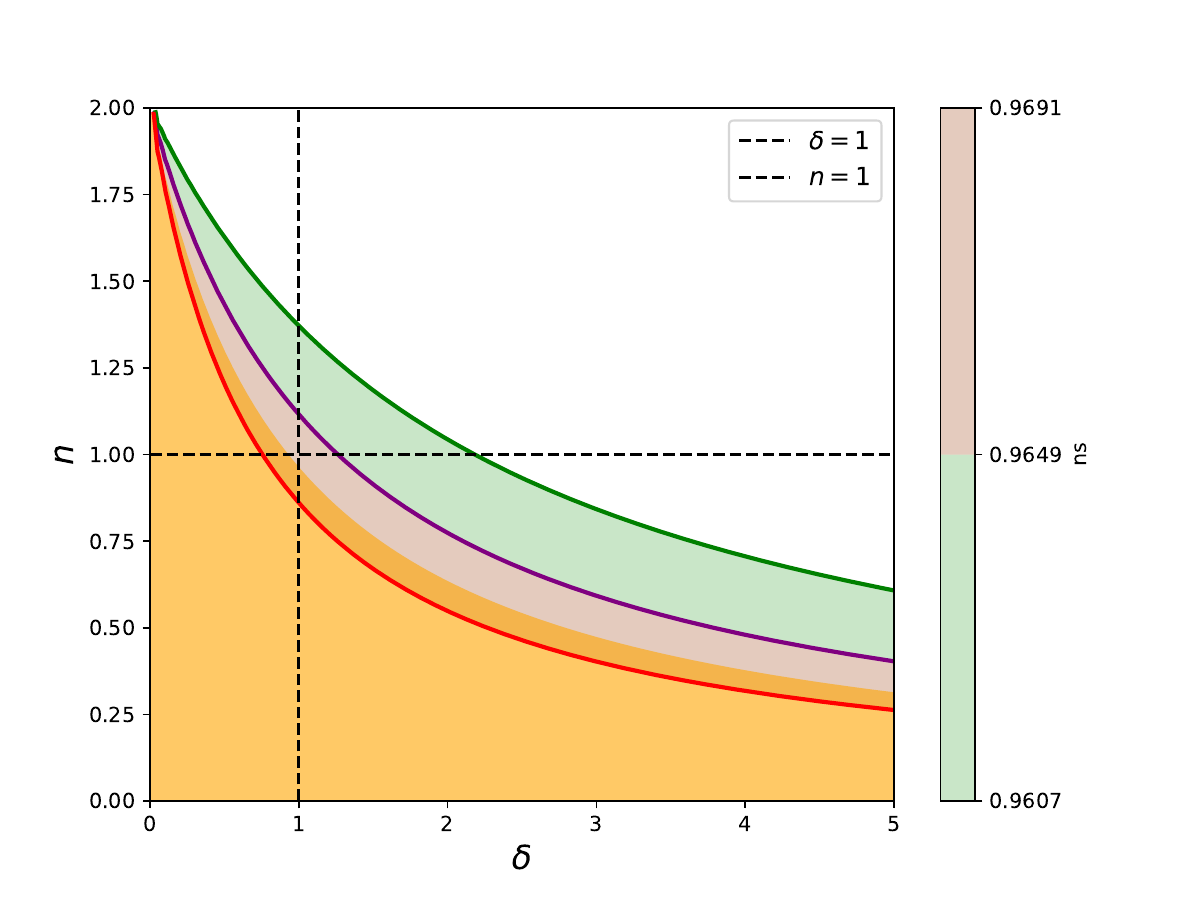}
		\caption{\small{Region of validity of the observable indices $n_s$ (purple) and $r$ (orange) given in Equation (\ref{nsr-tsal}) of Tsallis entropy
				with respect to the Planck data in an $n-\delta$ space. Here we take $N_f=60$}}
		\label{fig1}
	\end{center}
\end{figure}

\begin{figure}[th]
	\begin{center}
		\includegraphics[scale=0.48]{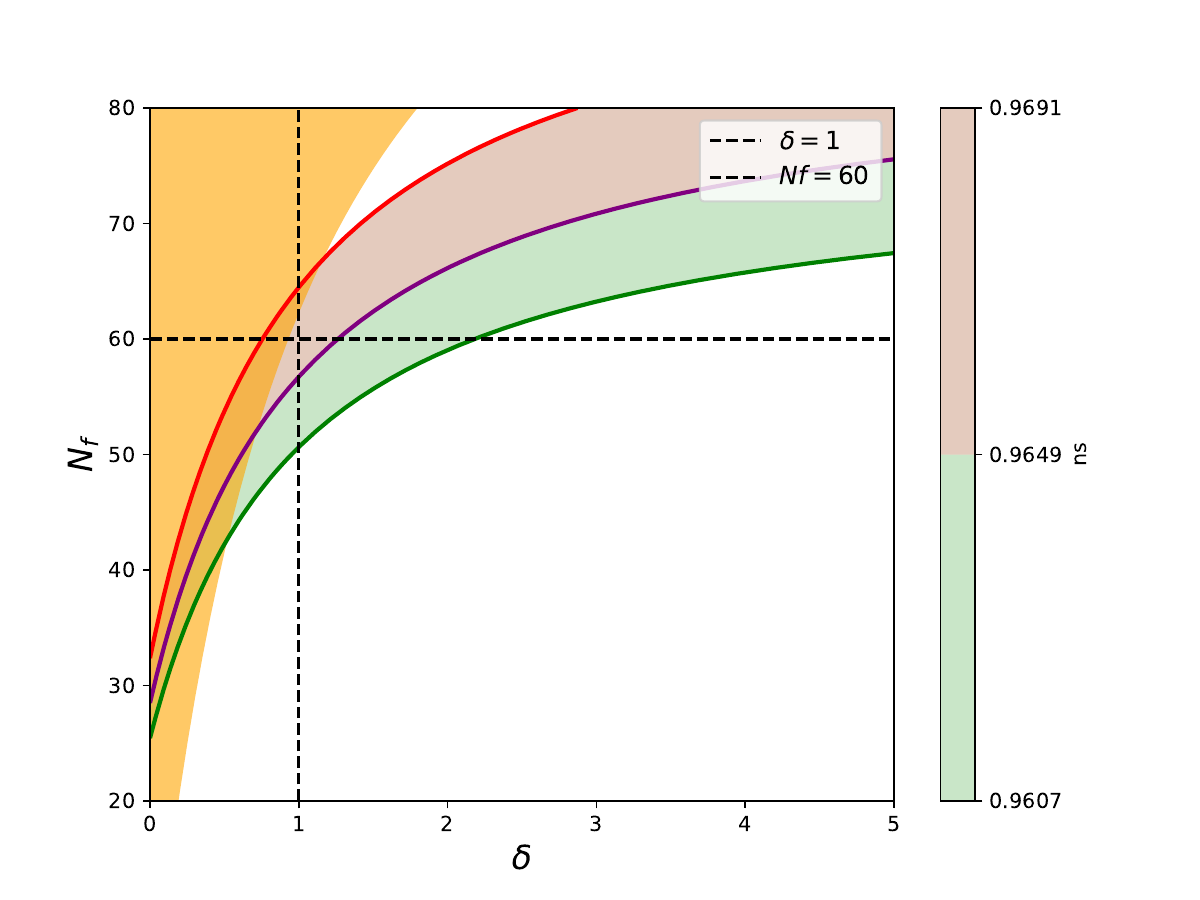}
		\caption{\small{Region of validity of the observable indices $n_s$ (purple) and $r$ (orange) given in Equation (\ref{nsr-tsal}) of Tsallis entropy
				with respect to the Planck data in an $N_f-\delta$ space. Here we take $n = 1$.}}
		\label{fig2}
	\end{center}
\end{figure}

From Eqs. (\ref{nsr-tsal}), Under Tsallis entropy, the observational indices $n_s$ and $r$ are influenced by the entropic parameter $\delta$, the number of e-folds $N_f$ and the power-law exponent $n$. In Fig. \ref{fig1}, under the Planck 2018 data (\ref{P2018}), the region of validity for two indices $\{n_s,~r\}$ is plotted in the $n-\delta$ space for a suitable e-folding $N_f=60$ (see Fig. \ref{fig1}).
It shows that the limiting case, $\delta \longrightarrow 0$, corresponds to $n \longrightarrow 2$. Additionally, in Fig. \ref{fig2}, the region of validity for $\{n_s,~r\}$  is depicted in the $N_f-n$ space for the case $n=1$, revealing a suitable region of $N_f$ around ($50\sim 60$) corresponds the parameter $\delta$ around  ($0.5 \sim 1$). Hence the power-law inflation for $n = 1$ up to $n = 2$ may yield viable inflationary outcomes aligning with Planck 2018 data when the apparent horizon has Tsallis entropy, contrasting with the BH entropy model's limitations.
\begin{figure}[th]
	\begin{center}
		\includegraphics[scale=0.48]{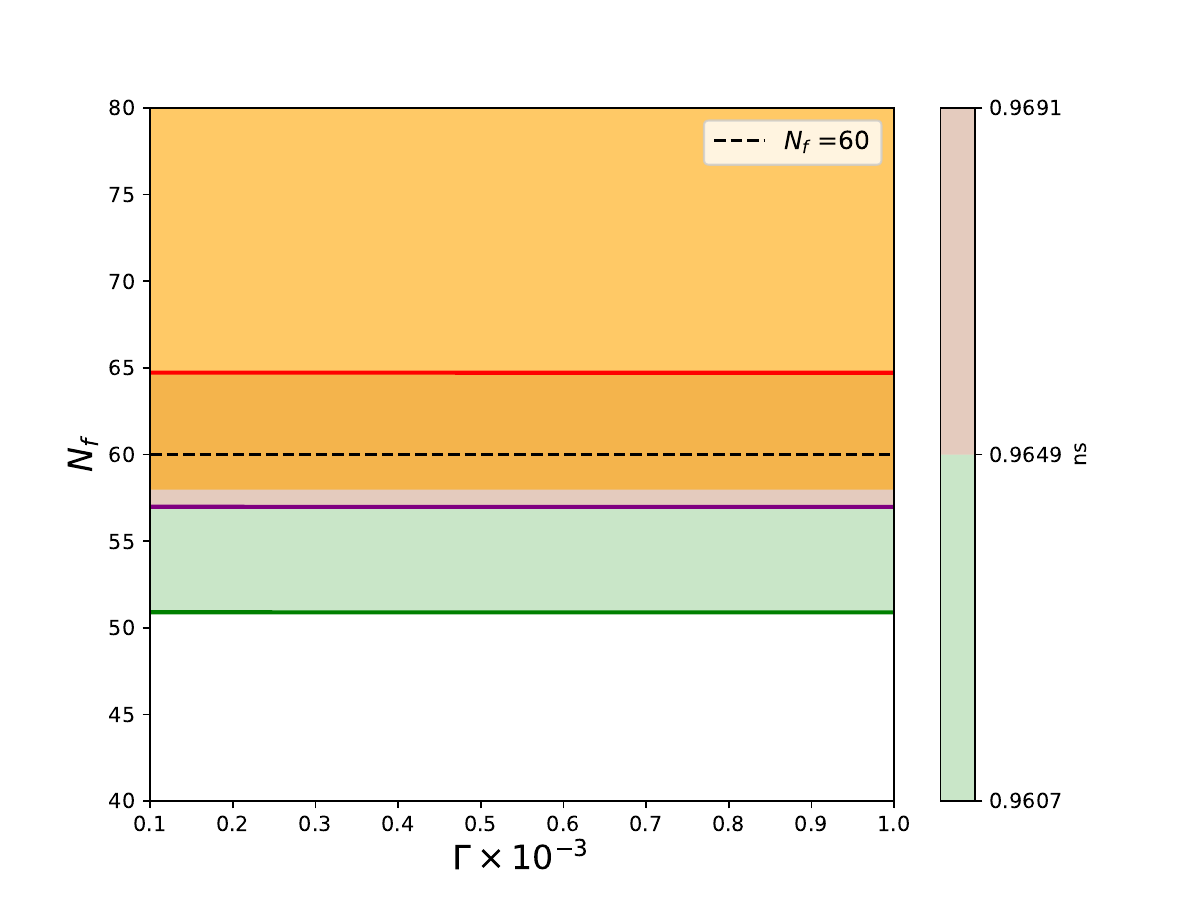}
		\caption{\small{Region of validity of the observable indices $n_s$ (purple) and $r$ (orange) given in Equation (\ref{nsr-Ren}) of R\'{e}nyi entropy
				with respect to the Planck data  in an $N_f-\Gamma$ space. Here we take $n = 1$.}}
		\label{fig3}
	\end{center}
\end{figure}
\begin{figure}[th]
\begin{center}
	\includegraphics[scale=0.48]{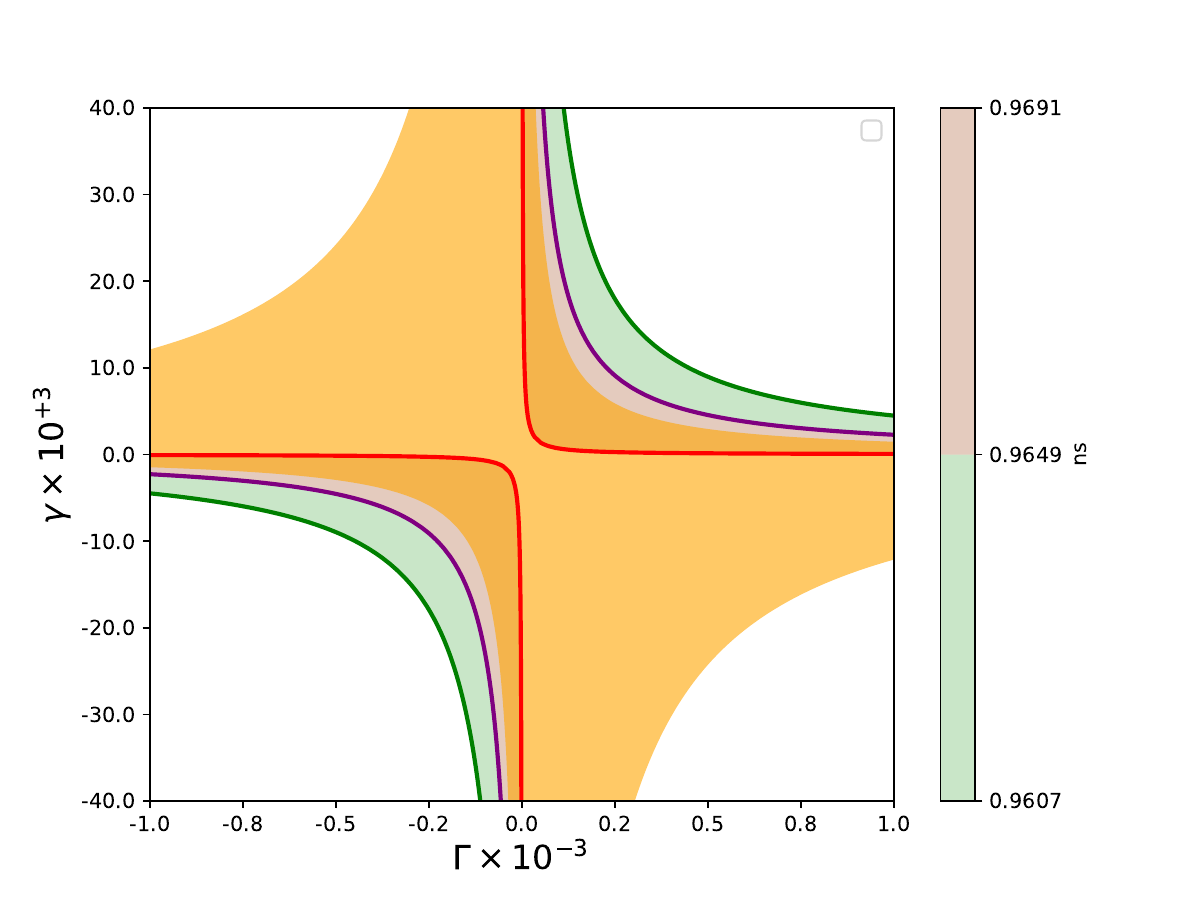}
	\caption{\small{Region of validity of the observable indices ns (purple) and r (orange) given in Equation (\ref{nsr-SM}) of Sharma-Mittal entropy
			with respect to the Planck data  in an $\gamma-\Gamma$ space. Here we take $n = 1$ and $N_f=65$.}}
	\label{fig4}
\end{center}
\end{figure}
\begin{figure}[th]
	\begin{center}
		\includegraphics[scale=0.48]{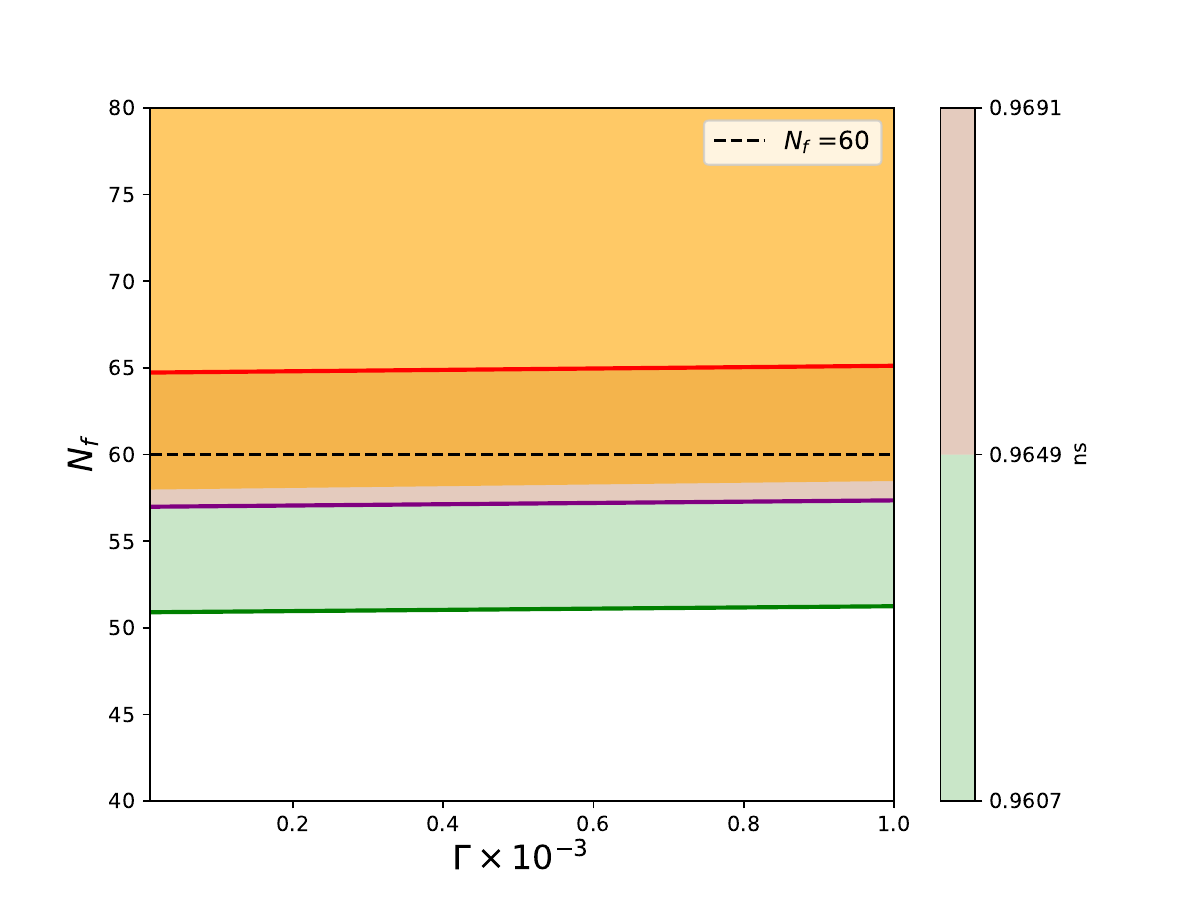}
		\caption{\small{Region of validity of the observable indices ns (purple) and r (orange) given in Equation (\ref{nsr-SM}) of Sharma-Mittal entropy
				with respect to the Planck data  in an $N_f-\Gamma$ space. Here we take $n = 1$ and $\gamma=100$.}}
		\label{fig5}
	\end{center}
\end{figure}
\subsection{R\'{e}nyi entropy}
From Eqs. (\ref{nsr-Ren}), Under R\'{e}nyi entropy, the observable indices $n_s$ and $r$ are influenced by the entropic parameter 
$\Gamma$ and the number of e-folds $N_f$.
It's notable that achieving perfect solutions for the indices for any $n$ is challenging under this framework, and the value $n=2$ leads to divergence in certain results. Consequently, our analysis is restricted to specific situations, particularly $n=1$. As illustrated in Fig. \ref{fig3}, the region of validity for $\{n_s,~r\}$ is plotted in the $N_f-\Gamma$ space for the case $n=1$, revealing a favorable region of $N_f$ around ($50\sim 60$) corresponds the parameter $\Gamma$ around  ($0.5 \sim 1$). Thus, the power-law potential inflation model for $n=1$ may offer a successful inflationary framework according to Planck data when utilizing R'{e}nyi entropy. In comparison with the Tsallis entropy case, this model suffers from inaccuracies and limitations of the power-law potential. However, it outperforms the conventional BH entropy model, demonstrating a higher potential for a successful inflationary model.
\subsection{Sharma-Mittal entropy}
Similar to the R\'{e}nyi entropy model, the Sharma-Mittal entropy model introduces an additional dimensionless parameter $\gamma$. However, similar challenges to those encountered in the R'{e}nyi entropy model arise when considering the Sharma-Mittal entropy model, particularly in the context of approximating the first order of $\Gamma$ and setting $n=1$. Despite these challenges, computational results suggest that the Sharma-Mittal entropy model outperforms the BH entropy model in explaining a viable power-law inflation scenario. Notably, an intriguing feature of the Sharma-Mittal entropy model is the freedom to assign both positive and negative values to $\Gamma$ and $\gamma$ in (\ref{nsr-SM}) simultaneously without encountering inconsistencies.  

 In Fig. \ref{fig4}, the region of validity for $\{n_s,~r\}$ is plotted in the $\gamma-\Gamma$ space for the case $n=1$ and suitable e-folding $N_f=65$. It demonstrates a wide range of overlap regions both in negative and positive values of parameters. Additionally, this validity regions are also plotted in the $N_f-\Gamma$ space in figure \ref{fig5}. As we see, two observational quantities $\{n_s,~r\}$ overlap in a wide range parameters with $\Gamma<10^{-3}$ and $N_f \ge 58$ for fixed parameters $n=1$ and typically $\gamma=100$. It is notable that for $\gamma=0$, the previous case (R\'{e}nyi entropy) will be retrieved. As a result we see that in this model, a viable power-law inflation may be achieved only for very small parameter values limited by $\alpha<10^{-8}$.

\section{Conclusion} \label{Sec5}
In this study, I explored the implications of incorporating non-extensive entropies within the framework of inflationary cosmology, focusing on a scalar field characterized by a power-law potential. My approach involved modifying the Friedmann equations by introducing generalized entropies that influence the energy-momentum tensor of the cosmic perfect fluid. I specifically examined the Tsallis, R\'{e}nyi, and Sharma-Mittal entropy forms. Notably, the introduction of a multiplicative factor in the energy-momentum density tensor, arising from the the non-extensive entropies, plays a crucial role in altering the evolution of the universe.

The results demonstrated that each entropy form leads to distinct modifications in inflationary dynamics, influencing the spectral index $n_s$ and tensor-to-scalar ratio $r$. Notably, the use of non-extensive entropies expands the parameter space for viable inflationary solutions, addressing limitations of the standard BH entropy framework. For each entropic model, I derived the slow-roll parameters, $\epsilon$ and $\eta$, which are critical for describing the inflationary phase. The calculations show that the slow-roll parameters are significantly influenced by the specific entropy model, impacting the spectral index $n_s$ and the tensor-to-scalar ratio $r$.

In the Tsallis entropy case, the modified Friedmann equations led to a distinct inflationary scenario with unique expressions for the slow-roll parameters. I demonstrated that the inflationary model is sensitive to the parameter $\delta$ of Tsallis entropy, with $n_s$ and $r$ aligning with observational constraints for specific values of $\delta$. Additionally, power-law inflation for $n = 1$ up to $n = 2$ may produce viable inflation with a graceful exit and align with the Planck 2018 data when considering Tsallis entropy.

In the R\'{e}nyi entropy case, I derived analytical expressions for the slow-roll parameters and showed how the extensively small parameter $\alpha$ or the consequently small value $\Gamma = \frac{3 \alpha}{8 V_0}$ of R\'{e}nyi entropy affects the inflationary predictions. The calculations and figure \ref{fig3}  showed that a viable cosmic inflation with a graceful exit is achieved for a wide range of the model parameter $\Gamma$, in a power-law potential with $n=1$ around the viable value of e-folding $N_f = 60$.

In the following, Sharma-Mittal entropy was studied. The slow-roll parameters versus entropic dimensionless parameters $\Gamma$ and $\gamma$ for $n=1$ were obtained. Similar to the R\'{e}nyi entropic case, it enables a wider range of overlap in the $n_s$ and $r$ space around e-folding $N_f = 65$. effectively accommodating observational data. As a result, in this model, viable power-law inflation with a graceful exit may be achieved only for very small parameter values limited by $\alpha < 10^{-8}$.

Calculations for models based on Kaniadakis, LQC and general four-parameters entropieswill involve successive approximations, which may cause some confusion and uncertainty in the results. In fact, as the complexity of the entropy model and the number of parameters increase, the accuracy of the results tends to decrease. From the point of view of computational difficulty, as demonstrated in Subsection \ref{Tsal2}, all quantities related to the calculation of $n_s$ and $r$ were determined with full accuracy and were dependent on the power index $n$. However, in Subsections \ref{Ren} and \ref{Sharm}, the calculations were restricted to the power index $n=1$ and were approximated only up to the first order of $\Gamma$. Consequently, the accuracy of the calculations in the Tsallis entropy case is higher. Therefore, when evaluating the suitability of the models under consideration, the Tsallis entropy model should be prioritized, followed by the R\'{e}nyi model, and lastly the Sharma-Mittal model, which has an additional parameter compared to the former. It is worth noting that the latter two models share common parameters of the same dimensions, which leads to greater similarities between them.

By comparing the results of non-extensive entropic models with those obtained from the standard BH entropy, it was shown that the non-extensive entropic models offer a broader range of inflationary solutions, resolving issues that arise in the standard framework. Specifically, while the BH entropic model typically provides a more constrained parameter space, the non-extensive entropies expand this space, offering more overlap with the observed $n_s$ and $r$ values.

In conclusion, this study underscores the potential of non-extensive entropies to enrich the theoretical framework of inflationary cosmology. By extending the conventional entropy paradigm, a deeper understanding of the early universe's dynamics is gained, aligning theoretical predictions with observational data. Future research could further explore the implications of these generalized entropies in various cosmological contexts, such as late-time acceleration and dark energy models.
\appendix
\section{Derivation of parameters $n_s$ and $r$ in the Presence of modified Friedmann Equation } \label{app}
In the presence of the correction function $f(\rho)$ in modified Fridmann equations (\ref{FR}), the modified slow-roll parameters (\ref{epseta}) are given by:
\begin{eqnarray}
\epsilon_{\text{mod}}& =& \frac{1}{16\pi} \left( \frac{V'}{V} \right)^2 \frac{1}{f(\rho) + \rho f'(\rho)} \notag \\
\eta_{\text{mod}}& =& \frac{1}{8\pi} \frac{V''}{V} \frac{1}{f(\rho)}.
\end{eqnarray}
The spectral index $n_s$ is related to the power spectrum of scalar perturbations by:
\begin{equation}
n_s - 1 = \frac{d\ln \mathcal{P}_\mathcal{R}}{d\ln k}.
\end{equation}
where the wavenumber $k=a H$ at the horizon crossing and the power spectrum of scalar perturbations $\mathcal{P}_\mathcal{R}$ is given by \cite{Mukhanov:1992, Liddle:2000}
\begin{equation}
\mathcal{P}_\mathcal{R} = \frac{H^2}{8\pi^2 M_{\text{pl}}^2\epsilon}.
\end{equation}
In the modified case, under the Planck mass unit, this becomes:
\begin{equation}
\mathcal{P}_\mathcal{R} = \frac{H^2}{\epsilon_{\text{mod}}}.
\end{equation}
Using the slow-roll approximation, this derivative can be expressed in terms of the modified $\epsilon$ and $\eta$ as
\begin{equation}
n_s - 1 \approx -6\epsilon_{\text{mod}} + 2\eta_{\text{mod}}.
\end{equation}
For the tensor to scalar ratio $r$, we must to consider the power spectrum of tensor perturbations $\mathcal{P}_t$ which in the standard case is given by \cite{Mukhanov:1992, Liddle:2000}: 
\begin{equation}
\mathcal{P}_t = \frac{2H^2}{\pi^2 M_{\text{pl}}^2}.
\end{equation}
In the modified case, the power spectrum of tensor perturbations remains unchanged because tensor perturbations depend only on $H$ and not on the form of the Friedmann equation. Therefore in Planck mass unit the tensor-to-scalar ratio $r$ gives
\begin{equation}
r = \frac{\mathcal{P}_t}{\mathcal{P}_\mathcal{R}}=16\epsilon_{\text{mod}}.
\end{equation}

In summary although modifying the Friedman equations changes the slow roll parameters of inflation, the basic relations of $n_s$ and $r$ remain unchanged in terms of $\epsilon$ and $\eta$.
\bibliography{refth} 

\begin{thebibliography}{43}
\expandafter\ifx\csname natexlab\endcsname\relax\def\natexlab#1{#1}\fi
\expandafter\ifx\csname bibnamefont\endcsname\relax
  \def\bibnamefont#1{#1}\fi
\expandafter\ifx\csname bibfnamefont\endcsname\relax
  \def\bibfnamefont#1{#1}\fi
\expandafter\ifx\csname citenamefont\endcsname\relax
  \def\citenamefont#1{#1}\fi
\expandafter\ifx\csname url\endcsname\relax
  \def\url#1{\texttt{#1}}\fi
\expandafter\ifx\csname urlprefix\endcsname\relax\def\urlprefix{URL }\fi
\providecommand{\bibinfo}[2]{#2}
\providecommand{\eprint}[2][]{\url{#2}}

\bibitem[{\citenamefont{Bekenstein}(1973)}]{Bekenstein:1973ur}
\bibinfo{author}{\bibfnamefont{J.~D.} \bibnamefont{Bekenstein}},
  \bibinfo{journal}{Phys. Rev. D} \textbf{\bibinfo{volume}{7}},
  \bibinfo{pages}{2333} (\bibinfo{year}{1973}).

\bibitem[{\citenamefont{Hawking}(1975)}]{Hawking:1975vcx}
\bibinfo{author}{\bibfnamefont{S.~W.} \bibnamefont{Hawking}},
  \bibinfo{journal}{Commun. Math. Phys.} \textbf{\bibinfo{volume}{43}},
  \bibinfo{pages}{199} (\bibinfo{year}{1975}), \bibinfo{note}{[Erratum:
  Commun.Math.Phys. 46, 206 (1976)]}.

\bibitem[{\citenamefont{Jacobson}(1995)}]{Jacobson:1995ab}
\bibinfo{author}{\bibfnamefont{T.}~\bibnamefont{Jacobson}},
  \bibinfo{journal}{Phys. Rev. Lett.} \textbf{\bibinfo{volume}{75}},
  \bibinfo{pages}{1260} (\bibinfo{year}{1995}), \eprint{gr-qc/9504004}.

\bibitem[{\citenamefont{Hayward}(1998)}]{Hayward:1997jp}
\bibinfo{author}{\bibfnamefont{S.~A.} \bibnamefont{Hayward}},
  \bibinfo{journal}{Class. Quant. Grav.} \textbf{\bibinfo{volume}{15}},
  \bibinfo{pages}{3147} (\bibinfo{year}{1998}), \eprint{gr-qc/9710089}.

\bibitem[{\citenamefont{Padmanabhan}(2005)}]{Padmanabhan:2003gd}
\bibinfo{author}{\bibfnamefont{T.}~\bibnamefont{Padmanabhan}},
  \bibinfo{journal}{Phys. Rept.} \textbf{\bibinfo{volume}{406}},
  \bibinfo{pages}{49} (\bibinfo{year}{2005}), \eprint{gr-qc/0311036}.

\bibitem[{\citenamefont{Padmanabhan}(2010)}]{Padmanabhan:2009vy}
\bibinfo{author}{\bibfnamefont{T.}~\bibnamefont{Padmanabhan}},
  \bibinfo{journal}{Rept. Prog. Phys.} \textbf{\bibinfo{volume}{73}},
  \bibinfo{pages}{046901} (\bibinfo{year}{2010}), \eprint{0911.5004}.

\bibitem[{\citenamefont{Padmanabhan}(2014)}]{Padmanabhan:2013nxa}
\bibinfo{author}{\bibfnamefont{T.}~\bibnamefont{Padmanabhan}},
  \bibinfo{journal}{Gen. Rel. Grav.} \textbf{\bibinfo{volume}{46}},
  \bibinfo{pages}{1673} (\bibinfo{year}{2014}), \eprint{1312.3253}.

\bibitem[{\citenamefont{Sanchez and Quevedo}(2023)}]{Sanchez:2022xfh}
\bibinfo{author}{\bibfnamefont{L.~M.} \bibnamefont{Sanchez}} \bibnamefont{and}
  \bibinfo{author}{\bibfnamefont{H.}~\bibnamefont{Quevedo}},
  \bibinfo{journal}{Phys. Lett. B} \textbf{\bibinfo{volume}{839}},
  \bibinfo{pages}{137778} (\bibinfo{year}{2023}), \eprint{2208.05729}.

\bibitem[{\citenamefont{Sheykhi}(2018)}]{Sheykhi_2018}
\bibinfo{author}{\bibfnamefont{A.}~\bibnamefont{Sheykhi}},
  \bibinfo{journal}{Physics Letters B} \textbf{\bibinfo{volume}{785}},
  \bibinfo{pages}{118–126} (\bibinfo{year}{2018}), ISSN
  \bibinfo{issn}{0370-2693},
  \urlprefix\url{http://dx.doi.org/10.1016/j.physletb.2018.08.036}.

\bibitem[{\citenamefont{Odintsov and Paul}(2023)}]{Odintsov:2022qnn}
\bibinfo{author}{\bibfnamefont{S.~D.} \bibnamefont{Odintsov}} \bibnamefont{and}
  \bibinfo{author}{\bibfnamefont{T.}~\bibnamefont{Paul}},
  \bibinfo{journal}{Phys. Dark Univ.} \textbf{\bibinfo{volume}{39}},
  \bibinfo{pages}{101159} (\bibinfo{year}{2023}), \eprint{2212.05531}.

\bibitem[{\citenamefont{Odintsov et~al.}(2023)\citenamefont{Odintsov,
  D’Onofrio, and Paul}}]{Odintsov_2023}
\bibinfo{author}{\bibfnamefont{S.~D.} \bibnamefont{Odintsov}},
  \bibinfo{author}{\bibfnamefont{S.}~\bibnamefont{D’Onofrio}},
  \bibnamefont{and} \bibinfo{author}{\bibfnamefont{T.}~\bibnamefont{Paul}},
  \bibinfo{journal}{Physics of the Dark Universe}
  \textbf{\bibinfo{volume}{42}}, \bibinfo{pages}{101277}
  (\bibinfo{year}{2023}), ISSN \bibinfo{issn}{2212-6864},
  \urlprefix\url{http://dx.doi.org/10.1016/j.dark.2023.101277}.

\bibitem[{\citenamefont{Odintsov et~al.}(2024)\citenamefont{Odintsov,
  D'Onofrio, and Paul}}]{Odintsov:2023rqf}
\bibinfo{author}{\bibfnamefont{S.~D.} \bibnamefont{Odintsov}},
  \bibinfo{author}{\bibfnamefont{S.}~\bibnamefont{D'Onofrio}},
  \bibnamefont{and} \bibinfo{author}{\bibfnamefont{T.}~\bibnamefont{Paul}},
  \bibinfo{journal}{Universe} \textbf{\bibinfo{volume}{10}}, \bibinfo{pages}{4}
  (\bibinfo{year}{2024}), \eprint{2312.13587}.

\bibitem[{\citenamefont{Lambiase et~al.}(2023)\citenamefont{Lambiase, Luciano,
  and Sheykhi}}]{Lambiase:2023ryq}
\bibinfo{author}{\bibfnamefont{G.}~\bibnamefont{Lambiase}},
  \bibinfo{author}{\bibfnamefont{G.~G.} \bibnamefont{Luciano}},
  \bibnamefont{and} \bibinfo{author}{\bibfnamefont{A.}~\bibnamefont{Sheykhi}},
  \bibinfo{journal}{Eur. Phys. J. C} \textbf{\bibinfo{volume}{83}},
  \bibinfo{pages}{936} (\bibinfo{year}{2023}), \eprint{2307.04027}.

\bibitem[{\citenamefont{Teimoori et~al.}(2024)\citenamefont{Teimoori,
  Rezazadeh, and Rostami}}]{Teimoori:2023hpv}
\bibinfo{author}{\bibfnamefont{Z.}~\bibnamefont{Teimoori}},
  \bibinfo{author}{\bibfnamefont{K.}~\bibnamefont{Rezazadeh}},
  \bibnamefont{and} \bibinfo{author}{\bibfnamefont{A.}~\bibnamefont{Rostami}},
  \bibinfo{journal}{Eur. Phys. J. C} \textbf{\bibinfo{volume}{84}},
  \bibinfo{pages}{80} (\bibinfo{year}{2024}), \eprint{2307.11437}.

\bibitem[{\citenamefont{Luciano}(2023)}]{Luciano:2023roh}
\bibinfo{author}{\bibfnamefont{G.~G.} \bibnamefont{Luciano}},
  \bibinfo{journal}{Eur. Phys. J. C} \textbf{\bibinfo{volume}{83}},
  \bibinfo{pages}{329} (\bibinfo{year}{2023}), \eprint{2301.12509}.

\bibitem[{\citenamefont{Brevik and Timoshkin}(2024)}]{Brevik:2024nzf}
\bibinfo{author}{\bibfnamefont{I.}~\bibnamefont{Brevik}} \bibnamefont{and}
  \bibinfo{author}{\bibfnamefont{A.~V.} \bibnamefont{Timoshkin}},
  \bibinfo{journal}{Int. J. Geom. Meth. Mod. Phys.}
  \textbf{\bibinfo{volume}{21}}, \bibinfo{pages}{2450181}
  (\bibinfo{year}{2024}), \eprint{2404.05597}.

\bibitem[{\citenamefont{Sheykhi and Ghaffari}(2023)}]{Sheykhi_2023}
\bibinfo{author}{\bibfnamefont{A.}~\bibnamefont{Sheykhi}} \bibnamefont{and}
  \bibinfo{author}{\bibfnamefont{S.}~\bibnamefont{Ghaffari}},
  \bibinfo{journal}{Physics of the Dark Universe}
  \textbf{\bibinfo{volume}{41}}, \bibinfo{pages}{101241}
  (\bibinfo{year}{2023}), ISSN \bibinfo{issn}{2212-6864},
  \urlprefix\url{http://dx.doi.org/10.1016/j.dark.2023.101241}.

\bibitem[{\citenamefont{P et~al.}(2023)\citenamefont{P, Pandey, Sharma, and
  Pankaj}}]{P:2022amn}
\bibinfo{author}{\bibfnamefont{S.~K.} \bibnamefont{P}},
  \bibinfo{author}{\bibfnamefont{B.~D.} \bibnamefont{Pandey}},
  \bibinfo{author}{\bibfnamefont{U.~K.} \bibnamefont{Sharma}},
  \bibnamefont{and} \bibinfo{author}{\bibnamefont{Pankaj}},
  \bibinfo{journal}{Eur. Phys. J. C} \textbf{\bibinfo{volume}{83}},
  \bibinfo{pages}{143} (\bibinfo{year}{2023}), \eprint{2211.15468}.

\bibitem[{\citenamefont{Nojiri et~al.}(2022)\citenamefont{Nojiri, Odintsov, and
  Faraoni}}]{Nojiri_2022}
\bibinfo{author}{\bibfnamefont{S.}~\bibnamefont{Nojiri}},
  \bibinfo{author}{\bibfnamefont{S.~D.} \bibnamefont{Odintsov}},
  \bibnamefont{and} \bibinfo{author}{\bibfnamefont{V.}~\bibnamefont{Faraoni}},
  \bibinfo{journal}{Astrophysics} \textbf{\bibinfo{volume}{65}},
  \bibinfo{pages}{534–551} (\bibinfo{year}{2022}), ISSN
  \bibinfo{issn}{1573-8191},
  \urlprefix\url{http://dx.doi.org/10.1007/s10511-023-09759-1}.

\bibitem[{\citenamefont{Manoharan et~al.}(2023)\citenamefont{Manoharan, Shaji,
  and Mathew}}]{Manoharan:2022qll}
\bibinfo{author}{\bibfnamefont{M.~T.} \bibnamefont{Manoharan}},
  \bibinfo{author}{\bibfnamefont{N.}~\bibnamefont{Shaji}}, \bibnamefont{and}
  \bibinfo{author}{\bibfnamefont{T.~K.} \bibnamefont{Mathew}},
  \bibinfo{journal}{Eur. Phys. J. C} \textbf{\bibinfo{volume}{83}},
  \bibinfo{pages}{19} (\bibinfo{year}{2023}), \eprint{2208.08736}.

\bibitem[{\citenamefont{Di~Gennaro and Ong}(2022)}]{Di_Gennaro_2022}
\bibinfo{author}{\bibfnamefont{S.}~\bibnamefont{Di~Gennaro}} \bibnamefont{and}
  \bibinfo{author}{\bibfnamefont{Y.~C.} \bibnamefont{Ong}},
  \bibinfo{journal}{Universe} \textbf{\bibinfo{volume}{8}},
  \bibinfo{pages}{541} (\bibinfo{year}{2022}), ISSN \bibinfo{issn}{2218-1997},
  \urlprefix\url{http://dx.doi.org/10.3390/universe8100541}.

\bibitem[{\citenamefont{Bhattacharjee}(2021)}]{Bhattacharjee_2021}
\bibinfo{author}{\bibfnamefont{S.}~\bibnamefont{Bhattacharjee}},
  \bibinfo{journal}{The European Physical Journal C}
  \textbf{\bibinfo{volume}{81}} (\bibinfo{year}{2021}), ISSN
  \bibinfo{issn}{1434-6052},
  \urlprefix\url{http://dx.doi.org/10.1140/epjc/s10052-021-09003-0}.

\bibitem[{\citenamefont{Moradpour et~al.}(2020)\citenamefont{Moradpour, Ziaie,
  and Kord~Zangeneh}}]{Moradpour:2020dfm}
\bibinfo{author}{\bibfnamefont{H.}~\bibnamefont{Moradpour}},
  \bibinfo{author}{\bibfnamefont{A.~H.} \bibnamefont{Ziaie}}, \bibnamefont{and}
  \bibinfo{author}{\bibfnamefont{M.}~\bibnamefont{Kord~Zangeneh}},
  \bibinfo{journal}{Eur. Phys. J. C} \textbf{\bibinfo{volume}{80}},
  \bibinfo{pages}{732} (\bibinfo{year}{2020}), \eprint{2005.06271}.

\bibitem[{\citenamefont{D'Agostino}(2019)}]{DAgostino:2019wko}
\bibinfo{author}{\bibfnamefont{R.}~\bibnamefont{D'Agostino}},
  \bibinfo{journal}{Phys. Rev. D} \textbf{\bibinfo{volume}{99}},
  \bibinfo{pages}{103524} (\bibinfo{year}{2019}), \eprint{1903.03836}.

\bibitem[{\citenamefont{Saridakis et~al.}(2018)\citenamefont{Saridakis, Bamba,
  Myrzakulov, and Anagnostopoulos}}]{Saridakis_2018}
\bibinfo{author}{\bibfnamefont{E.~N.} \bibnamefont{Saridakis}},
  \bibinfo{author}{\bibfnamefont{K.}~\bibnamefont{Bamba}},
  \bibinfo{author}{\bibfnamefont{R.}~\bibnamefont{Myrzakulov}},
  \bibnamefont{and} \bibinfo{author}{\bibfnamefont{F.~K.}
  \bibnamefont{Anagnostopoulos}}, \bibinfo{journal}{Journal of Cosmology and
  Astroparticle Physics} \textbf{\bibinfo{volume}{2018}},
  \bibinfo{pages}{012–012} (\bibinfo{year}{2018}), ISSN
  \bibinfo{issn}{1475-7516},
  \urlprefix\url{http://dx.doi.org/10.1088/1475-7516/2018/12/012}.

\bibitem[{\citenamefont{Hernández-Almada
  et~al.}(2022)\citenamefont{Hernández-Almada, Leon, Magaña,
  García-Aspeitia, Motta, Saridakis, Yesmakhanova, and
  Millano}}]{Hern_ndez_Almada_2022}
\bibinfo{author}{\bibfnamefont{A.}~\bibnamefont{Hernández-Almada}},
  \bibinfo{author}{\bibfnamefont{G.}~\bibnamefont{Leon}},
  \bibinfo{author}{\bibfnamefont{J.}~\bibnamefont{Magaña}},
  \bibinfo{author}{\bibfnamefont{M.~A.} \bibnamefont{García-Aspeitia}},
  \bibinfo{author}{\bibfnamefont{V.}~\bibnamefont{Motta}},
  \bibinfo{author}{\bibfnamefont{E.~N.} \bibnamefont{Saridakis}},
  \bibinfo{author}{\bibfnamefont{K.}~\bibnamefont{Yesmakhanova}},
  \bibnamefont{and} \bibinfo{author}{\bibfnamefont{A.~D.}
  \bibnamefont{Millano}}, \bibinfo{journal}{Monthly Notices of the Royal
  Astronomical Society} \textbf{\bibinfo{volume}{512}},
  \bibinfo{pages}{5122–5134} (\bibinfo{year}{2022}), ISSN
  \bibinfo{issn}{1365-2966},
  \urlprefix\url{http://dx.doi.org/10.1093/mnras/stac795}.

\bibitem[{\citenamefont{Asghari and Sheykhi}(2022)}]{Asghari:2021bqa}
\bibinfo{author}{\bibfnamefont{M.}~\bibnamefont{Asghari}} \bibnamefont{and}
  \bibinfo{author}{\bibfnamefont{A.}~\bibnamefont{Sheykhi}},
  \bibinfo{journal}{Eur. Phys. J. C} \textbf{\bibinfo{volume}{82}},
  \bibinfo{pages}{388} (\bibinfo{year}{2022}), \eprint{2110.00059}.

\bibitem[{\citenamefont{Dabrowski and Salzano}(2020)}]{Dabrowski:2020atl}
\bibinfo{author}{\bibfnamefont{M.~P.} \bibnamefont{Dabrowski}}
  \bibnamefont{and} \bibinfo{author}{\bibfnamefont{V.}~\bibnamefont{Salzano}},
  \bibinfo{journal}{Phys. Rev. D} \textbf{\bibinfo{volume}{102}},
  \bibinfo{pages}{064047} (\bibinfo{year}{2020}), \eprint{2009.08306}.

\bibitem[{\citenamefont{Tsallis}(1988)}]{Tsallis:1987eu}
\bibinfo{author}{\bibfnamefont{C.}~\bibnamefont{Tsallis}}, \bibinfo{journal}{J.
  Statist. Phys.} \textbf{\bibinfo{volume}{52}}, \bibinfo{pages}{479}
  (\bibinfo{year}{1988}).

\bibitem[{\citenamefont{R\'{e}nyi}(1960)}]{Renyi:1960}
\bibinfo{author}{\bibfnamefont{A.}~\bibnamefont{R\'{e}nyi}},
  \bibinfo{journal}{Proceedings of the 4th Berkeley Symposium on Mathematics,
  Statistics and Probability, University of California Press, Berkeley and Los
  Angeles} \textbf{\bibinfo{volume}{I}}, \bibinfo{pages}{547}
  (\bibinfo{year}{1960}).

\bibitem[{\citenamefont{Sayahian~Jahromi
  et~al.}(2018)\citenamefont{Sayahian~Jahromi, Moosavi, Moradpour,
  Morais~Gra\c{c}a, Lobo, Salako, and Jawad}}]{SayahianJahromi:2018irq}
\bibinfo{author}{\bibfnamefont{A.}~\bibnamefont{Sayahian~Jahromi}},
  \bibinfo{author}{\bibfnamefont{S.~A.} \bibnamefont{Moosavi}},
  \bibinfo{author}{\bibfnamefont{H.}~\bibnamefont{Moradpour}},
  \bibinfo{author}{\bibfnamefont{J.~P.} \bibnamefont{Morais~Gra\c{c}a}},
  \bibinfo{author}{\bibfnamefont{I.~P.} \bibnamefont{Lobo}},
  \bibinfo{author}{\bibfnamefont{I.~G.} \bibnamefont{Salako}},
  \bibnamefont{and} \bibinfo{author}{\bibfnamefont{A.}~\bibnamefont{Jawad}},
  \bibinfo{journal}{Phys. Lett. B} \textbf{\bibinfo{volume}{780}},
  \bibinfo{pages}{21} (\bibinfo{year}{2018}), \eprint{1802.07722}.

\bibitem[{\citenamefont{Barrow}(2020)}]{Barrow:2020tzx}
\bibinfo{author}{\bibfnamefont{J.~D.} \bibnamefont{Barrow}},
  \bibinfo{journal}{Phys. Lett. B} \textbf{\bibinfo{volume}{808}},
  \bibinfo{pages}{135643} (\bibinfo{year}{2020}), \eprint{2004.09444}.

\bibitem[{\citenamefont{Kaniadakis}(2005)}]{Kaniadakis:2005zk}
\bibinfo{author}{\bibfnamefont{G.}~\bibnamefont{Kaniadakis}},
  \bibinfo{journal}{Phys. Rev. E} \textbf{\bibinfo{volume}{72}},
  \bibinfo{pages}{036108} (\bibinfo{year}{2005}), \eprint{cond-mat/0507311}.

\bibitem[{\citenamefont{Drepanou et~al.}(2022)\citenamefont{Drepanou, Lymperis,
  Saridakis, and Yesmakhanova}}]{Drepanou:2021jiv}
\bibinfo{author}{\bibfnamefont{N.}~\bibnamefont{Drepanou}},
  \bibinfo{author}{\bibfnamefont{A.}~\bibnamefont{Lymperis}},
  \bibinfo{author}{\bibfnamefont{E.~N.} \bibnamefont{Saridakis}},
  \bibnamefont{and}
  \bibinfo{author}{\bibfnamefont{K.}~\bibnamefont{Yesmakhanova}},
  \bibinfo{journal}{Eur. Phys. J. C} \textbf{\bibinfo{volume}{82}},
  \bibinfo{pages}{449} (\bibinfo{year}{2022}), \eprint{2109.09181}.

\bibitem[{\citenamefont{Majhi}(2017)}]{Majhi:2017zao}
\bibinfo{author}{\bibfnamefont{A.}~\bibnamefont{Majhi}},
  \bibinfo{journal}{Phys. Lett. B} \textbf{\bibinfo{volume}{775}},
  \bibinfo{pages}{32} (\bibinfo{year}{2017}), \eprint{1703.09355}.

\bibitem[{\citenamefont{Liu}(2022)}]{Liu:2021dvj}
\bibinfo{author}{\bibfnamefont{Y.}~\bibnamefont{Liu}}, \bibinfo{journal}{EPL}
  \textbf{\bibinfo{volume}{138}}, \bibinfo{pages}{39001}
  (\bibinfo{year}{2022}), \eprint{2112.15077}.

\bibitem[{\citenamefont{Khodam-Mohammadi and
  Monshizadeh}(2023)}]{Khodam_Mohammadi_2023}
\bibinfo{author}{\bibfnamefont{A.}~\bibnamefont{Khodam-Mohammadi}}
  \bibnamefont{and}
  \bibinfo{author}{\bibfnamefont{M.}~\bibnamefont{Monshizadeh}},
  \bibinfo{journal}{Physics Letters B} \textbf{\bibinfo{volume}{843}},
  \bibinfo{pages}{138066} (\bibinfo{year}{2023}), ISSN
  \bibinfo{issn}{0370-2693},
  \urlprefix\url{http://dx.doi.org/10.1016/j.physletb.2023.138066}.

\bibitem[{\citenamefont{Cai and Kim}(2005)}]{Cai:2005ra}
\bibinfo{author}{\bibfnamefont{R.-G.} \bibnamefont{Cai}} \bibnamefont{and}
  \bibinfo{author}{\bibfnamefont{S.~P.} \bibnamefont{Kim}},
  \bibinfo{journal}{JHEP} \textbf{\bibinfo{volume}{02}}, \bibinfo{pages}{050}
  (\bibinfo{year}{2005}), \eprint{hep-th/0501055}.

\bibitem[{\citenamefont{Bin\'etruy and Helou}(2015)}]{Binetruy:2014ela}
\bibinfo{author}{\bibfnamefont{P.}~\bibnamefont{Bin\'etruy}} \bibnamefont{and}
  \bibinfo{author}{\bibfnamefont{A.}~\bibnamefont{Helou}},
  \bibinfo{journal}{Class. Quant. Grav.} \textbf{\bibinfo{volume}{32}},
  \bibinfo{pages}{205006} (\bibinfo{year}{2015}), \eprint{1406.1658}.

\bibitem[{\citenamefont{Mukhanov}(2005)}]{mukhanov2005physical}
\bibinfo{author}{\bibfnamefont{V.}~\bibnamefont{Mukhanov}},
  \emph{\bibinfo{title}{Physical Foundations of Cosmology}}
  (\bibinfo{publisher}{Cambridge University Press},
  \bibinfo{address}{Cambridge, UK}, \bibinfo{year}{2005}), ISBN
  \bibinfo{isbn}{978-0-521-56398-7}.

\bibitem[{\citenamefont{Tsallis}(2009)}]{tsallis2009introduction}
\bibinfo{author}{\bibfnamefont{C.}~\bibnamefont{Tsallis}},
  \emph{\bibinfo{title}{Introduction to Nonextensive Statistical Mechanics:
  Approaching a Complex World}} (\bibinfo{publisher}{Springer},
  \bibinfo{address}{New York, NY}, \bibinfo{year}{2009}), ISBN
  \bibinfo{isbn}{978-0-387-85358-1}.

\bibitem[{\citenamefont{Mukhanov et~al.}(1992)\citenamefont{Mukhanov, Feldman,
  and Brandenberger}}]{Mukhanov:1992}
\bibinfo{author}{\bibfnamefont{V.~F.} \bibnamefont{Mukhanov}},
  \bibinfo{author}{\bibfnamefont{H.~A.} \bibnamefont{Feldman}},
  \bibnamefont{and} \bibinfo{author}{\bibfnamefont{R.~H.}
  \bibnamefont{Brandenberger}}, \bibinfo{journal}{Physics Reports}
  \textbf{\bibinfo{volume}{215}}, \bibinfo{pages}{203} (\bibinfo{year}{1992}).

\bibitem[{\citenamefont{Liddle and Lyth}(2000)}]{Liddle:2000}
\bibinfo{author}{\bibfnamefont{A.~R.} \bibnamefont{Liddle}} \bibnamefont{and}
  \bibinfo{author}{\bibfnamefont{D.~H.} \bibnamefont{Lyth}},
  \emph{\bibinfo{title}{{Cosmological Inflation and Large-Scale Structure}}}
  (\bibinfo{publisher}{Cambridge University Press},
  \bibinfo{address}{Cambridge, UK}, \bibinfo{year}{2000}), ISBN
  \bibinfo{isbn}{978-0-521-57598-0}.

\end{thebibliography}
\end{document}